\documentstyle[12pt]{article}
\newcommand{\ra}{\rightarrow}

\newcommand{\ee}{$e^+e^-$\ }

\newcommand{\beq}{\begin{eqnarray*}}
\newcommand{\eeq}{\end{eqnarray*}}

\newcommand{\SUSY}{{\cal SUSY}}
\newcommand{\MSSM}{{\cal MSSM}}

\newcommand{\pp} {$pp$\mbox{ }}
\newbox\mycount
\newcommand{\ctowidth}[2]{ \setbox\mycount=\hbox{$#2$}
                          \hbox to \wd\mycount{$ \hss #1 \hss $} }
\newcommand{\ltowidth}[2]{ \setbox\mycount=\hbox{$#2$}
                          \hbox to \wd\mycount{$\hskip0pt plus0pt minus1fil
                           #1 \hfill $} }
\newcommand{\rtowidth}[2]{ \setbox\mycount=\hbox{$#2$}
                          \hbox to \wd\mycount{$\hfill #1
                          \hskip0pt plus0pt minus1fil$} }
 
\begin{document}
 
\begin{titlepage}
 
\begin{flushright}
DESY 93--173\\
UCD--93--39\\
December 1993 \\
\end{flushright}
 
\vspace{1cm}
 
\begin{center}
 
{\large\sc $\gamma \gamma$ PRODUCTION OF NON--STRONGLY
INTERACTING\\ [5mm]
$\cal{SUSY}$ PARTICLES AT HADRON COLLIDERS}
 
\vspace{1cm}
 
{\large J.\ Ohnemus$^1$, T.F.\ Walsh$^2$, and P.M.\ Zerwas$^3$\\
\vspace{1cm}
\noindent
$^1$ Physics Department, Univ. of California, Davis, CA 95616, USA\\
$^2$ Physics Dept., Univ. of Minnesota, Minneapolis, MN 55455, USA\\
$^3$ Deutsches Elektronensynchrotron DESY, D-22603 Hamburg, FRG\\
}
\end{center}
 
\vspace{5cm}
 
\begin{abstract}
\noindent
Non--strongly interacting supersymmetric particles -- sleptons,
charginos, neutral\-inos, and charged Higgs bosons -- are difficult to
detect at the Large Hadron Collider. We therefore examine the
possibility of producing particles of this type in virtual $\gamma
\gamma$ collisions at the LHC. Since photons can be emitted from
protons which do not break up in the radiation process, very clean
events can be generated, compensating to some extent for the small
event numbers. Higher rates are expected, at the expense of stray
hadrons, for events in which one or two protons break up.
\end{abstract}
 
\end{titlepage}
 
\section[]{$\gamma \gamma$ Luminosities}
 
Colored supersymmetric particles -- squarks and gluinos -- will be
produced copiously at the Large Hadron Collider (LHC) \cite{1}. The
mass range up to about 2~TeV can be investigated in this sector.
Non--strongly interacting supersymmetric particles -- sleptons,
charginos/neutralinos, and charged Higgs bosons -- can be easily
detected at \ee colliders. However, it is much more challenging to
find these particles at hadron colliders. The Drell--Yan and $gg$
fusion mechanisms yield low production rates for these particles,
futhermore, the processes of interest are embedded in a complicated,
jet-filled environment. Charged Higgs bosons can be observed as decay
products of top quarks if their mass is sufficiently small. Cascade
decays of squarks and gluinos can lead to final states which include
charginos and neutralinos. However, the analysis of cascades
containing several types of unknown particles is not a simple
experimental task.  Thus one may wonder whether other reactions could
be exploited to supplement these experimental searches.
 
In this note we examine the possibility of searching for non--strongly
interacting $\SUSY$ particles in $\gamma \gamma$ collisions at hadron
colliders. The disadvantage of this method is obviously the low
$\gamma \gamma$ luminosity, suppressed essentially by two powers of
$\alpha$ and counterbalanced only partly by large logarithmic
enhancement factors. However, the disadvantage of the low production
rates is compensated to some extent by the simple topology of the
initial state and the potentially clean environment of the final
state. If the high energy photons are emitted without breaking up the
initial protons in the radiation process or by exciting low-mass
states, the situation is experimentally simple. Since the protons or
the low mass hadron beam fragments continue travelling in the beam
directions, a clean virtual photon beam will be generated. The
transverse momentum of the photon is cut off by the radius of the
proton (or some other smaller but fixed radius for low mass fragments)
so that the photon  spectrum is scale--invariant.
 
The $\gamma \gamma$ luminosity based on the strictly elastic channel
$pp \rightarrow pp + \gamma \gamma$ is shown by the curve labeled [$e
\ast e$] in Fig.~1. For small photon energies the spectrum of photons
emitted from protons is given by
\begin{equation}
   f_ {\gamma/p} (x) = \frac{\alpha}{\pi} \, \frac{1}{x}
\, {\rm log} \left( \frac{r_E^2}{x^2} \right) \>,
\end{equation}
where $r^2_E = 0.71\ {\rm GeV}^2 / M^2_p$ and $M_p$ is the proton
mass. The photon radiation is dominated by the electric monopole
transition. This is the standard Weizs\"acker-Williams spectrum for
photons with the virtuality integrated between the minimum value
$|t_{min}| \approx 0.71\ {\rm GeV}^2$ and the maximum value $|t_{max}|
= M^2_p x^2/(1-x) $ defined by the charge radius of the proton
\cite{2}. The corresponding $\gamma \gamma$ luminosity, normalized to
the $pp$ luminosity, can be written approximately in leading order of
${\rm log} \, (1/ \tau)$ as
\begin{equation}
  \tau \, \frac{d {\cal L}^{ee}}{d \tau}
 = \left( \frac{\alpha}{\pi} \right) ^2 \, \frac{2}{3} \, {\rm log}^3
 \left( \frac{1}{\tau} \right) \>.
\end{equation}
These are minimal values of the luminosity since additional
low--resonance excitations will raise the photon yield.
 
Besides the coherent emission of photons from the protons, photons can
also be radiated from the quarks and antiquarks in processes involving
the break--up of the protons, $pp \rightarrow XX + \gamma \gamma$. In
this case the photons are accompanied by spectator hadrons. The
corresponding $\gamma \gamma$ luminosities, based on the leading order
GRV parametrization of the parton densities \cite{3}, are displayed by
the curve labeled [$i \ast i$] for the LHC with $\sqrt{s}= 14$~GeV in
Fig.~1. For small values of $ \tau $ the $\gamma \gamma$ luminosity
may  be roughly approximated by
\begin{equation}
 \tau \frac{d {\cal L}^{ii}}{d \tau} = \frac{\rho^2}{120}
 \, \left( \frac{\alpha}{\pi} \right)^2 \, {\rm log}^2
\left( \frac{\tau s}{M^2_c} \right) \,
{\rm log}^5  \left( \frac{1}{\tau} \right) \>,
\end{equation}
where the proton structure function has been crudely parametrized as $
F_2 \approx \rho \, {\rm log} \, (1/x)$ with $\rho = 0.16$ and $x$
\raisebox{-.6ex}{$\stackrel{>}{\sim}$} $10^{-3}$. $Q^2 \approx \tau s$
is the typical scale of the $\gamma \gamma$ subprocess and the
parameter $M_c$ may be chosen as 1~GeV, the effective non-perturbative
transverse momentum cut-off. (Since the formula is restricted to
leading logarithmic accuracy, the precise numerical value of this
quantity is inessential.) The copious number of $q \overline{q}$ sea
pairs at high energies leads to luminosities significantly larger than
the elastic [$e \ast e$] channel.
 
By combining photons emitted from protons with photons emitted from
quarks, a third class of $\gamma \gamma$ events is generated in \pp
collisions, $pp \rightarrow pX + \gamma \gamma$. The corresponding
luminosities are shown by the curve labeled [$e \ast i$] in Fig.~1. In
leading logarithmic order, with the same approximations as before, the
mixed luminosity reads
\begin{equation}
 \tau \frac{d {\cal L}^{ei}}{d \tau} = \frac{\rho}{6}
 \, \left( \frac{\alpha}{\pi} \right)^2 \, {\rm log}
\left( \frac{\tau s}{M^2_c} \right) \,
{\rm log}^4  \left( \frac{1}{\tau} \right) \>.
\end{equation}
In this class of events, spectator hadrons spray only into one
hemisphere, while a pure $\gamma$ beam, unaccompanied by hadronic
debris, is generated in the opposite hemisphere.
 
\vspace{-0.5cm}
 
\section[]{$\cal{SUSY}$ Processes}
 
{\it (i)} \underline{\it Charginos} couple to photons in the standard
way of spin 1/2 fermions. Including the mass effects at the threshold,
the cross section \cite{3A} for producing a pair of identical
charginos is
\begin{equation}
\sigma (\gamma \gamma \ra {\tilde{\chi}}^+ {\tilde{\chi}}^-) 
= \frac{4 \pi \alpha^2}{s} \,
\left [ \left ( 1 + \frac
{4 m^2}{s} - \frac{8 m^4}{s^2} \right ) \,
{\rm log} \left( \frac{1 + \beta}
{1 - \beta} \right) - \beta\, 
\left (1 + \frac{4 m^2}{s} \right ) \right ] \>,
\end{equation}
where $\sqrt{s}$ is the $\gamma \gamma$ c.m. energy, $\beta = (1 - 4
m^2 /s)^{\frac{1}{2}}$ is the velocity of the charginos in the c.m.
frame, and $m$ is the chargino mass. In the minimal supersymmetric
extension of the Standard Model [$\MSSM$], there are two chargino
states, which are mixtures of the winos and charged higgsinos. The
mass of the lighter of these states could be of $\cal O$(100 GeV),
while the heavier state may have mass of $\cal O$(300 GeV).
 
{\it (ii)} \underline{\it Sleptons} are spin--zero particles, the
cross section for producing them is
\begin{eqnarray}
\sigma (\gamma \gamma \ra \tilde{\ell}^+ \tilde{\ell}^-)
= \frac{2 \pi \alpha^2}{s} \, \left [\,\beta \, \left (1 + \frac
{4 m^2}{s} \right ) - \frac{4 m^2}{s} \left (1 - \frac{2 m^2}
{s} \right ) \, {\rm log}
\left( \frac{1 + \beta}{1 - \beta} \right) \, \right ] \>.
\end{eqnarray}
The cross section is not logarithmically enhanced, thus the production
rate for sleptons is expected to be significantly lower than the
chargino rate. There are two slepton states in the $\MSSM$ for each
family species, related to the left-- and right--handed lepton
components.
 
{\it (iii)} \underline{\it Charged Higgs bosons}, $\gamma \gamma \ra
H^+ H^-$, will be produced at the same rate as sleptons in $\gamma
\gamma$ collisions.
 
The cross sections for the elastic--elastic $\gamma \gamma$ production
of sleptons, charged Higgs bosons, and charginos are plotted versus
the mass of the produced particle (denoted generically by $m$) in
Fig.~2a for the LHC ($\sqrt{s}$ = 14~TeV). These cross sections are
for events in which the protons do not break up. Including one-sided
elastic events in which only one of the protons does not break up,
will increase the particle yield by about a factor of 5 as shown in
Fig.~2b. Similarly, if both protons break up, the cross section
increases by an additional $5 \sim 10\%$ as shown in Fig.~2c. Given an
integrated luminosity of $10\ {\rm fb}^{-1}$ at the LHC for one year
of operation, about 10 chargino pairs will be generated for a chargino
mass of 150 GeV in elastic--elastic events; if one-sided elastic
events are also included, this number increases to about 50 chargino
pairs. Given the clean final states, these event numbers appear large
enough to be interesting. Slepton and charged Higgs boson cross
sections are smaller by about a factor 5. For an integrated luminosity
of $100\ {\rm fb}^{-1}$ per year a sample of 500 chargino pairs will
be generated at a mass of 150 GeV.\footnote{This agrees with the
corresponding estimates for heavy lepton rates of $e \ast e$ and $i
\ast i$ type processes at the LHC in Ref.~\cite{3B}.} Even though a
pile-up of events will occur in this situation, the very small
transverse momenta of the pile-up events will help to eliminate them.
exceed the yield that could be expected at dedicated photon--proton
colliders in which invariant energies $\sqrt{s_{\gamma p}}$ of 1 TeV
may be reached \cite{4}. However, they are significantly smaller than
the event rates at $\gamma \gamma$ colliders in which the photons are
generated by Compton back-scattering of laser light at $e^+e^-$ linear
colliders.
 
Among other decay modes, charginos decay into $W$'s and the lightest
neutralino (LSP) which is assumed to be stable in the present context.
Sleptons may decay into their leptonic Standard Model partners and the
LSP, charged Higgs bosons decay into $t \overline{b}$ etc. Therefore
we have also calculated the background cross sections for $\gamma
\gamma \ra W^+ W^-$ and $\gamma \gamma \ra \mu^+ \mu^-$. Since these
cross sections become very large for forward/backward scattering, we
have applied cuts on the rapidity, $ \left| y \right| \leq 2$, and the
transverse momentum, $p_T^{} >$ 20 GeV, of the final state particles.
The invariant mass of the pair is also required to satisfy $M > 2\,
m$. Despite these cuts, the $W^+ W^-$ and $\mu^+ \mu^-$ backgrounds
are still much larger than the signal cross sections. However, in
$\gamma \gamma \ra W^+ W^-$ and $\gamma \gamma \ra \mu^+ \mu^-$ events
the transverse momenta of the $W$'s or $\mu$'s are balanced, in
contrast to the signal events in which the transverse momenta of the
final state $W$ or $\mu$ particles are different due to the escaping
invisible LSP's. By rejecting events in which the $W$ or $\mu$
transverse momenta balance to within about 10~GeV, the backgrounds are
reduced considerably while the signals are little affected. 

{\it In summa}. $\gamma \gamma$ collisions induced by initial state
bremsstrahlung of photons at the LHC may provide production channels
to search for non--strongly interacting supersymmetric particles. The
low rates are compensated in part by the clean environment in which
the novel particles can be generated in the class of two- and
one-sided elastic $\gamma \gamma$ events. These events can in
principle be selected by vetoing events with visible forward particles
in both hemispheres.
 
The final state topology of doubly inelastic $\gamma \gamma$ events is
somewhat similar (but not identical) to the Drell-Yan processes, which
have larger cross sections. Without a detailed calculation including
the distributions for the two processes and the detector acceptance,
it is not possible to make a clear statement on how useful the $\gamma
\gamma$ process will prove to be in this case. We prefer to emphasize
here the clean nature of the doubly elastic and elastic-inelastic
events.
 
Of course, we cannot analyze the experimental  triggering problems for
the elastic-elastic and elastic-inelastic events. Our present
theoretical estimates are intended as an exploratory survey of the 
general physics opportunities in $\gamma \gamma$ collisions at the
LHC.
  
\section*{\large \bf Acknowledgments}
 
JO would like to thank the DESY theory group for hospitality
during the initial stages of this work.
 
\newpage
 
\section*{\large \bf Figures}
 
\begin{enumerate}
\item Luminosities for $\gamma \gamma$ collisions at the LHC
($\sqrt{s}=14$~TeV). Luminosities are shown for the elastic--elastic
[$e \ast e$], elastic--inelastic [$e \ast i$], and
inelastic--inelastic [$i \ast i$] processes.
\item Cross sections for chargino, slepton, and charged Higgs boson
production in $\gamma \gamma$ collisions at the LHC
($\sqrt{s}=14$~TeV). The cross sections are plotted versus the mass of
the produced particle which is denoted generically by $m$. Also shown
are the cross sections for the background processes
$\gamma \gamma \to W^+ W^-$ and $\gamma \gamma \to \mu^+ \mu^-$, with
the cuts described in the text. Parts a), b), and c) are for the
elastic--elastic, elastic--inelastic, and inelastic--inelastic
processes, respectively.
\end{enumerate}

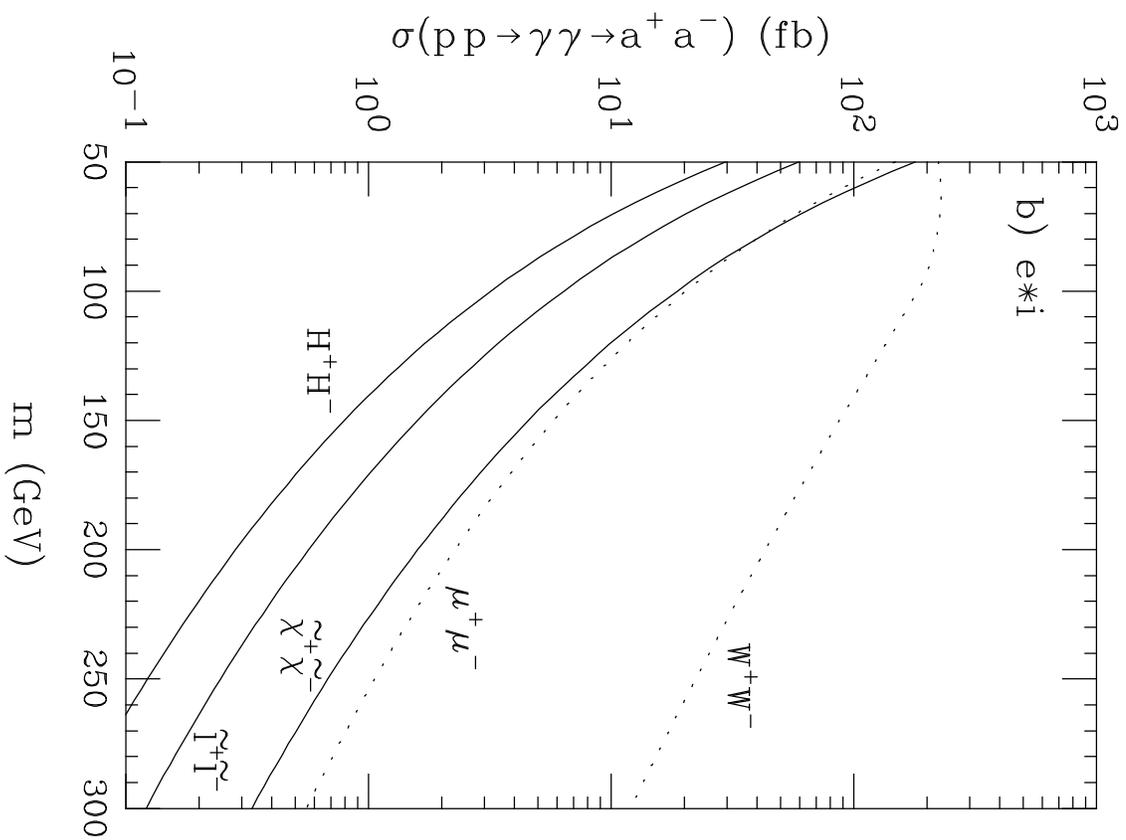
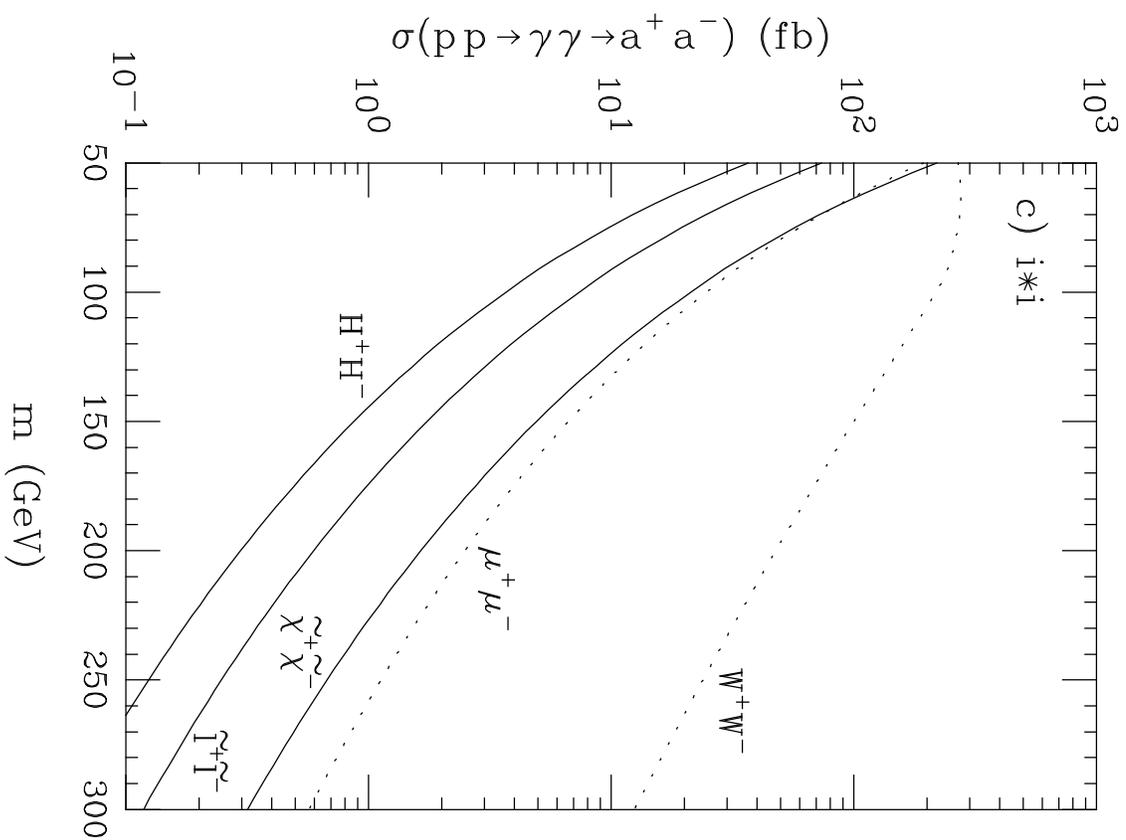
\end{document}